\documentclass[preprint,showpacs,showkeys,preprintnumbers,amsmath,amssymb,aps]{revtex4-1}

\usepackage{bm,amssymb,amsmath,mathrsfs}
\usepackage[usenames]{color}
\usepackage{hyperref}

\usepackage{dcolumn}
\usepackage[pdftex]{graphicx}

\newcommand{\bc}{\begin{center}}
\newcommand{\ec}{\end{center}}
\newcommand{\bit}{\begin{itemize}}
\newcommand{\eit}{\end{itemize}}
\newcommand{\bq}{\begin{equation}}
\newcommand{\eq}{\end{equation}}
\newcommand{\pprime}{{\prime\prime}}

\begin{document}

\title{Polarization at TLEP/FCC-ee: ideas and estimates}

\author{S.~R.~Mane}
\email{srmane001@gmail.com}

\affiliation{Convergent Computing Inc., P.~O.~Box 561, Shoreham, NY 11786, USA}

\begin{abstract}
I examine various ideas for radiative polarization at TLEP/FCC-ee 
and formulate some estimates for the polarization buildup time and the asymptotic polarization.
Using wigglers, a useful degree of polarization (for energy calibration), 
with a time constant of about 1 h, 
may be possible up to the threshold of $W$ pair production.
At higher energies such as the threshold of Higgs production, 
attaining a useful level of polarization may be difficult in a planar ring.
With Siberian Snakes, wigglers and some imagination, polarization of reasonable magnitude, 
with a reasonable time constant (of not more than about 1 h), may be achievable at very high energies. 
\end{abstract}



\maketitle

\baselineskip 3ex
\tableofcontents

\baselineskip 4ex

\vfill\pagebreak
\section{Introduction}
There is interest in polarized beams in very high energy circular $e^+e^-$ colliders,
both for energy calibration and for longitudinally polarized colliding beams.
I shall discuss both topics below.
TLEP has been renamed FCC-ee, but for contact with work by others I shall continue to use `TLEP' occasionally. 

A study of polarization in TLEP was presented by Uli Wienands \cite{Wienands2013}
in April 2013 at the Snowmass 2013 Lepton Collider Workshop at MIT
(a run-up meeting for ``Snowmass on the Mississippi'').
Slide 8 of Wienands's presentation displays a graph of the estimated asymptotic polarization in TLEP
over an energy interval from 40 to 180 GeV.
The graph of the asymptotic polarization diplays a curious shape:
it is flat at $P/P_{ST} = 0.6$ for energies $E = 40-80$ GeV and drops steeply at higher energies $E > 80$ GeV.
(Here $P_{ST} = 8/(5/\sqrt3) \simeq 92.4\%$ is the Sokolov-Ternov polarization.)
For ease of reference, I sketch a similar looking graph in Fig.~\ref{fig:figulislide8} below. 
(Note that there is no physics content to my sketch in Fig.~\ref{fig:figulislide8},
I simply made up a similar looking curve for ease of reference. The plotted function is
$P/P_{ST} = 0.07 + (0.53/2)[1+\tanh((95-E)/7)] -0.02\exp\{-(E-140)^2/40^2\}$.)
Wienands \cite{Wienandsprivcomm} kindly informed me that he employed a simple model of
quantum excitation and damping to model the spin diffusion and depolarization at high energies (above 80 GeV).
However, the asymptotic polarization was cut off at $P/P_{ST} = 0.6$ for $E \le 80$ GeV and there is no explanation for this.

\vskip 0.2in
I present my own analysis of the polarization in TLEP/FCC-ee below.
Note that the FCC-ee ring design is still very flexible, so I shall try to keep my analysis to a general level.
For general information on spin dynamics in accelerators, I direct the reader to the reviews 
by Yuri Shatunov, Kaoru Yokoya and myself \cite{MSY1,MSY2}.

\subsection*{\em Disclaimer}
{\em 
I am not affiliated with any university or research laboratory.
The analysis below is entirely my own, taken on my own initiative.
Having said this, I have benefitted from communications with various colleagues,
who I do not name, for obvious reasons.
}

\vfill\pagebreak
\section{Sokolov-Ternov polarization time}
\subsection{Basic results}
The first step is to calculate the Sokolov-Ternov polarization time,
for a flat ring without wigglers, depolarizing resonances or other complications.
I employ the parameter values from \cite{TLEPparams}, which lists the current machine parameters for FCC-ee.
There are four operating energies of principal interest, viz.~the $Z^0$ pole (45 GeV),
$W$-pair production threshold (80 GeV),
Higgs production (technically $e^+e^- \to HZ$, 120 GeV)
and $t\bar{t}$ pair production (175 GeV).
The polarization time for an isomagnetic ring of circumference $2\pi R$ and bend radius $\rho$ is given by
\bq
\tau_p = 98.66\,\frac{\rho^3}{E^5}\,\frac{R}{\rho} \,.
\eq
The time is in seconds, the lengths are in m and the energy is in GeV.
I tabulate the corresponding Sokolov-Ternov polarization times (in minutes) in Table \ref{tab:ring_params}, 
together with some results for LEP.
The columns are chosen to match the table in \cite{TLEPparams}.

For a fixed ring circumference and bend radius, the Sokolov-Ternov polarization time varies inversely as the fifth power of the beam energy.
Hence for FCC-ee at the $Z^0$ pole, the Sokolov-Ternov polarization time is about 270 h,
whereas at 120 GeV (Higgs) it is about 2h and at 175 GeV ($t\bar{t}$) it is only 19 min.
This shows that a speedup mechanism is required at 45 GeV.
The only serious candidate to speed up the polarization time is to employ asymmetric wigglers,
but wigglers increase the energy spread and the strengths of the depolarizing spin resonances.
All of this has long been known.
I shall quantify the use of wigglers more carefully below.

\vfill\pagebreak
\subsection{Top-up mode and luminosity lifetime}
FCC-ee will employ full energy injection.
The bunches will have a relatively short luminosity lifetime (see \cite{TLEPparams} and Table \ref{tab:ring_params}) 
and will be continuously topped up.
Under these circumstances, the polarization will not grow exponentially, starting from zero.
The freshly injected electrons (or positrons) will be unpolarized, and we must compute a `weighted average' polarization.
If the luminosity lifetime is $\tau_\ell$, then with a top-up rate so that the steady-state bunch population is $N_0$, we have
\bq
\frac{dN}{dt} = -\frac{N}{\tau_\ell} +\frac{N_0}{\tau_\ell} \,.
\eq
I shall use the term `lifetime weighted polarization' and it is given by
\bq
\label{eq:avgpol}
\begin{split}
P_{\rm avg} &= \frac{1}{N_0}\int_0^\infty \frac{N_0}{\tau_\ell}\,e^{-t/\tau_\ell}\,P_0(1-e^{-t/\tau_p})\, dt
\\
&= P_0\,\biggl[\, \int_0^\infty \frac{e^{-t/\tau_\ell}}{\tau_\ell}\, dt -\int_0^\infty \frac{e^{-t(1/\tau_\ell+1/\tau_p)}}{\tau_\ell}\, dt \,\biggr]
\\
&= P_0\,\biggl[\, 1 - \frac{\tau_p}{\tau_\ell+\tau_p} \,\biggr] 
\\
&= P_0\,\frac{\tau_\ell}{\tau_\ell+\tau_p} 
= \frac{P_0}{1 + (\tau_p/\tau_\ell)} \,.
\end{split}
\eq
If $\tau_\ell=\infty$ (or $\tau_\ell\gg\tau_p$),
which is the usual approximation in polarization calculations,
then the asymptotic polarization equals $P_0$.
However, if $\tau_p = \tau_\ell$, the average polarization is a factor of 2 smaller than $P_0$.
If $\tau_\ell\ll\tau_p$ (luminosity lifetime much shorter than polarization time),
the average polarization can be much less than $P_0$.

\subsection{Pilot bunches (non-colliding)}
Another possibility in FCC-ee is to circulate a few `pilot bunches' which do not collide with an opposing beam.
These non-colliding bunches are not topped up and have a longer circulation lifetime in the ring.
Hence their polarization buildup time can be longer than 1 h.
These non-colliding bunches will be employed for energy calibration only.
I shall discuss their use below.

\vfill\pagebreak
\section{Asymmetric wigglers}
At a beam energy of 45 GeV, the guide field is quite low
\bq
B = \frac{B\rho}{\rho} = \frac{45 \times 3.3356}{11000} \simeq 0.0136 \;\textrm{T} = 136 \;\textrm{Gauss}\,.
\eq
The magnetic guide field at other energies is tabulated in Table \ref{tab:ring_params}. 
Even at 175 GeV, the guide field in FCC-ee is comparable to that in LEP at 45 GeV.
Hence it should be possible to build asymmetric wigglers of any desired magnetic field, 
to speed up the polarization time to any desired value.
The constraints on the wigglers will arise from the increase in the radiated power, 
and the increase in the beam energy spread,
and the concomitant increase in the strengths of the depolarizing spin resonances.
The radiation loss per turn is proportional to the synchrotron radiation integral
\bq
I_2 = \oint \frac{ds}{\rho^2} \,.
\eq
The Sokolov-Ternov polarization rate is proportional to the synchrotron radiation integral
\bq
I_3 = \oint \frac{ds}{|\rho|^3} \,.
\eq
For later use, for the numerator term for the asymptotic polarization, let me define
\bq
I_{3a} = \oint \frac{\hat{\bm{b}}\cdot\hat{\bm{n}}_0}{|\rho|^3}\,ds \,.
\eq
The speedup factor of the polarization rate is $A_p \propto I_3$, 
the ratio of the increase in the radiation loss per turn is $A_U \propto I_2$
and the ratio of the increase in the relative beam energy spread is $A_e = \sqrt{I_3/I_2}$
({\em but see below}).
Let me also define a ratio $A_n$ for the numerator of the formula for the asymptotic polarization.
If the value of $A_U$ exceeds unity, then to constrain the radiated power at 100 MW (= $2 \times 50$ MW per beam),
it will be necessary to decrease the single beam current by a factor of $A_U$.
This tradeoff must be borne in mind as a constraint on the wiggler parameters.

I shall treat only asymmetric wigglers consisting of a triplet with bend radii in the ratios $-1:\frac12:-1$.
The Sokolov-Ternov asymptotic polarization, from the wigglers only, is
\bq
\frac{P}{P_{ST}} = \frac{-1+8-1}{1+8+1} = \frac{6}{10} = 0.6 \,.
\eq
If the bend dipoles in FCC-ee are 10 m long, the bend angle per dipole is $\theta_b = 10/11000$ rad.
The total number of bend dipoles is $2\pi/\theta_b = 2\pi\times1100 \simeq 6911$.
For simplicity, I shall assume the ring has $10^4$ arc dipoles.
I shall also assume there are 100 wigglers in the ring.
Define a ratio $\mathcal{N}$ such that
\bq
\frac{1}{|\rho|_{\rm wiggler}} = \frac{\mathcal{N}}{|\rho|_{\rm arc}} \,.
\eq
Then the synchrotron radiation integrals change in the ratios
\begin{subequations}
\begin{align}
A_U &= 1+ \frac{100}{10000}\,\mathcal{N}^2
&=& 1 + 0.01\, \mathcal{N}^2 \,, 
\\
A_p &= 1 + \frac{100}{10000}\,\mathcal{N}^3 
&=& 1 + 0.01\, \mathcal{N}^3 \,, 
\\
A_n &= 1 + 0.6\,\frac{100}{10000}\,\mathcal{N}^3
&=& 1 + 0.006\, \mathcal{N}^3 \,, 
\\
A_e &= \sqrt{\frac{1 + 0.01\, \mathcal{N}^3}{1 + 0.01\, \mathcal{N}^2}} \,.
\end{align}
\end{subequations}
The value of $\mathcal{N}$ will depend on the beam energy. 
For now, consider $E=45$ GeV and $\mathcal{N} = 30$. 
This requires an easily achievable wiggler field.
Then
\bq
A_U = 10 \,,\qquad
A_p = 271 \,,\qquad
A_n = 163 \,,\qquad
A_e \simeq 5.2 \,,\qquad
\frac{P_{wig}}{P_{ST}} = \frac{A_n}{A_p} \simeq 0.601 \,.
\eq
At a beam energy of 45.5 GeV, where $\tau_p \simeq 270$ h, 
this will na{\"\i}vely speed up the polarization time to approximately 1h.
This analysis of course neglects the contribution of spin resonances, which I shall treat later.
The single beam current must be decreased by a factor of 10, to avoid increasing the radiated power.
The relative energy spread will increase by a factor of about 5.
From the information in \cite{TLEPparams}, 
the relative energy spread at 45.5 GeV is $0.06\%$, so this will increase the value to $0.3\%$.

\vfill\pagebreak
\section{Energy spread}
The data for the relative energy spread in \cite{TLEPparams} 
indicates a nonnegligible contribution from beamstrahlung (BS).
For example at 45.5 GeV in FCC-ee, 
the relative energy spread due to synchrotron radiation (SR) is $0.04\%$,
but with beamstrahlung included, it is $0.06\%$.
We can write
\bq
\sigma_\varepsilon^2 = \sigma_{\varepsilon,\,SR}^2 + \sigma_{\varepsilon,\,BS}^2 \,. 
\eq
Here $\sigma_{\varepsilon,\,SR}$ is the value without wigglers
and $\sigma_{\varepsilon,\,BS}$ is the contribution from beamstrahlung.
The energy spread due to the wigglers increases only the contribution from the synchrotron radiation.
We must write
\bq
\sigma_\varepsilon^2 = \frac{I_3}{I_2}\,\sigma_{\varepsilon,\,SR}^2 + \sigma_{\varepsilon,\,BS}^2 \,. 
\eq

\vfill\pagebreak
\section{Spin resonances}
\subsection{General remarks}
The analysis of polarization in very high energy $e^+e^-$ storage rings 
is incomplete without inclusion of the effects of spin resonances.
Here is where the data from the LEP Energy Model will prove invaluable.
LEP was the ring where a large body of data for the polarization was compiled,
at multiple energies all in the same ring,
which demonstrated the behavior of the asymptotic polarization (and the time constant),
in particular the important situation where the higher order spin resonances
began to dominate over the first order spin resonances.
Such data simply did not exist before LEP.
Furthermore, `polarization wigglers' were used for the first time at LEP.
They did {\em not} help:
the increased energy spread made the spin resonances too strong,
reducing the asymptotic polarization to an unacceptably small value.
(However, the polarization wigglers were useful for other aspects of LEP operations.)
This is all valuable information.
We shall need to leverage the experience gained from the LEP Energy Model.
\bit
\item
LEP I and II together compiled data on the asymptotic polarization over an energy interval from 40 to 100 GeV
(see, e.g.~\cite{LEP_EPAC1994} and \cite{LEP_review}).
This was the first time that polarization data had been compiled over a large energy interval, all in the same ring.
\item
HERA demonstrated the successful implementation of so-called `strong spin matching,' 
which is essential to attain longitudinally polarized colliding beams.
This is important, but is a separate issue.
\item
The polarization was also measured at SPEAR \cite{Spear_pol}, over a much smaller energy interval,
but with maps of numerous first and higher order spin resonances.
The most important feature of the SPEAR data, in the present context, was the 
synchrotron sideband resonances of a parent first order (horizontal) betatron resonance. 
Both Jean Buon \cite{Buon_1990} and I \cite{Mane_SPEAR} fitted the widths of the 
synchrotron sideband resonances and demonstrated
that the values from the analytical theoretical formulas were in agreement with the data.
Note that the analytical formulas had {\em no} adjustable parameters, so this was a zero-parameter fit,
which is an important validation of the theory to be used below.
\eit
In addition, the team of the LEP Energy Model also plotted graphs of the maximum asymptotic polarization 
attained at numerous $e^+e^-$ colliders, from VEPP-2M and ACO upwards.
See Fig.~3 in \cite{LEP_EPAC1994} (data up to 1994)
and Fig.~10 in \cite{LEP_review} (data up to 2000).
A significant lesson to learn from these plots
is that the maximum asymptotic polarization followed a simple scaling law,
based on first order resonances only, for a surprisingly large energy range.
The polarization was parameterized via (eq.~1 in \cite{LEP_EPAC1994})
\bq
\label{eq:polscale1}
P = \frac{P_{ST}}{1 + (\alpha E)^2} \,.
\eq
Here $\alpha$ is a phenomenological parameter.
The maximum asymptotic polarization attained at numerous $e^+e^-$ colliders
were all fitted by the above curve, with a single value of $\alpha$.
(Technically, one value of $\alpha$ for the data without harmonic spin matching,
and a different value of $\alpha$ for the data with harmonic spin matching.)
The effects of higher order spin resonances became significant at higher energies above the $Z^0$ pole
and the polarization decreased more rapidly than the power law in eq.~\eqref{eq:polscale1}.
To model the higher order spin resonances, I include a so-called
`depolarization enhancement factor' $\mathcal{F}$ into eq.~\eqref{eq:polscale1}.
I shall explain below how to calculate the depolarization enhancement factor.
Then
\bq
\label{eq:polscale2}
P = \frac{P_{ST}}{1 + (\alpha E)^2\mathcal{F}} \,.
\eq
In Fig.~\ref{fig:pol_scaling},
I display graphs of the asymptotic polarization using the above scaling law.
The dashed curves were computed using eq.~\eqref{eq:polscale1}
and the solid curves were computed using eq.~\eqref{eq:polscale2}.
I display graphs to simulate both the cases with and without harmonic spin matching (HSM).
I computed the curves in Fig.~\ref{fig:pol_scaling} using the following values for $\alpha$
\bq
\label{eq:alpha_LEP}
\alpha = \begin{cases} 0.06 &\qquad \textrm{without HSM} \\
0.015 & \qquad \textrm{with HSM} \,.
\end{cases}
\eq
The curves (very) approximately match the data shown in Fig.~3 in \cite{LEP_EPAC1994} and Fig.~10 in \cite{LEP_review}.
Harmonic spin matching will of course be essential at FCC-ee.
I shall argue below that a reasonable value to employ for FCC-ee is
to decrease the above HSM value of $\alpha$ by about $2$, so
\bq
\label{eq:alpha_TLEP}
\alpha \simeq 0.007 \qquad \textrm{(FCC-ee, with HSM)} \,.
\eq

\vfill\pagebreak
\subsection{Driving terms of spin resonances}
As the experience of the LEP Energy Model demonstrated, 
the synchrotron sideband spin resonances were by far the most spin resonances.
However, by definition, {\em sidebands} require a parent resonance.
I shall argue that synchrotron sideband spin resonances centered on an integer
are the most important set of spin resonances.
The most notable paper for the synchrotron sideband resonances
was published in 1983 by Kaoru Yokoya \cite{Yokoya_syncsdb_1983}.
Yokoya gave expressions (so-called `spin integrals') 
for the strengths of the first order spin resonances
in terms of one-turn integrals of lattice functions,
while the strengths of the synchrotron sideband resonances were expressed 
using formulas involving modified Bessel functions.
(Technically, the spin integrals for the first order spin resonances were known before 1983.)
The expressions for the sideband resonances were encapsulated into
so-called  `depolarization enhancement factors' 
which multiplied the strengths of the parent spin resonances.
I follow the treatment in \cite{Yokoya_syncsdb_1983}.
The spin precession equation can be expressed as
\bq
\frac{d\bm{s}}{d\theta} = (\bm{\Omega}_0+\bm{\omega}) \times \bm{s} \,.
\eq
Here $\bm{\Omega}_0$ is the spin precession vector on the (imperfect) closed orbit,
and $\bm{\omega}$ describes the perturbation due to the off-axis orbital motion. 
The spin motion on the closed orbit can be expressed using a right-handed orthonormal triad
$(\bm{l}_0,\bm{m}_0,\bm{n}_0)$,
where $\bm{n}_0$ is periodic around the ring 
and $\bm{l}_0$ and $\bm{m}_0$ precess around $\bm{n}_0$ at the spin tune $\nu$.
Then define $\bm{k}_0=\bm{l}_0+i\bm{m}_0$ and parameterize $\bm{n}$ via
$\bm{n} = \sqrt{1-|\zeta|^2}\,\bm{n}_0 +\Re(\bm{k}_0^*\zeta)$.
The strengths of the spin resonances are given by 
$\partial\zeta/\partial\varepsilon$, 
where $\varepsilon = \Delta E/E_0 = \Delta\gamma/\gamma_0$
and $\zeta$ satisfies the equation
\bq
\label{eq:dzetadtheta}
\frac{d\zeta}{d\theta} = -i\bm{\omega}\cdot\bm{k}_0 \sqrt{1-|\zeta|^2}
+i\bm{\omega}\cdot\bm{n}_0\,\zeta \,.
\eq
For first order spin resonances 
\bq
\label{eq:zeta1}
\zeta = -i \int\bm{\omega}\cdot\bm{k}_0 \,d\theta^\prime \,.
\eq
For the higher order spin resonances, it is complicated to integrate eq.~\eqref{eq:dzetadtheta},
but in general the most important driving term for $\zeta$ is given by
(eq.~(2.14) in \cite{Yokoya_syncsdb_1983})
\bq
\label{eq:zeta2}
\zeta = -i e^{-i\chi(\theta)} \int e^{i\chi(\theta^\prime)} \, \bm{\omega}\cdot\bm{k}_0\, d\theta^\prime \,,\qquad
\chi = -\int\bm{\omega}\cdot\bm{n}_0 \,d\theta^\pprime \,.
\eq
Hence we need to analyze the structure of 
$\int\bm{\omega}\cdot\bm{k}_0\,d\theta^\prime$ and
$\int\bm{\omega}\cdot\bm{n}_0\,d\theta^\prime$.
Following Yokoya \cite{Yokoya_syncsdb_1983}, 
we decompose $\bm{\omega}$ into the contributions from the various orbital modes
(eq.~(2.2) in \cite{Yokoya_syncsdb_1983})
\bq
\bm{\omega} = \bm{\omega}_\varepsilon\,\varepsilon +\bm{\omega}_x\,x_\beta +\bm{\omega}_y\,y_\beta \,.
\eq
The notation is self-explanatory.
For ultrarelativistic particles (eqs.~(2.3)--(2.5) in \cite{Yokoya_syncsdb_1983})
\begin{align}
\bm{\omega}_\varepsilon &= 
-R\,\Bigl[\,(\gamma_0a+1)G_x\eta_x -\frac{1}{\rho_x}\,\Bigr]\,\bm{e}_y
+R\,\Bigl[\,(\gamma_0a+1)G_y\eta_y -\frac{1}{\rho_y}\,\Bigr]\,\bm{e}_x \,,
\\
\bm{\omega}_x &= -R(\gamma_0a+1)G_x\,\bm{e}_y \,,
\\
\bm{\omega}_y &= \phantom{-}R(\gamma_0a+1)G_y\,\bm{e}_x \,.
\end{align}
The above expressions neglect fringe field terms, which is a good approximation for $\gamma_0a\gg1$.
Here $G_{x,y}$ are the quadrupole focusing gradients and $\rho_{x,y}$ are the bend radii.
Also $\eta_{x,y}$ are the horizontal and vertical dispersion functions.
Typically $1/\rho_y = \eta_y=0$ in the absence of lattice imperfections.
Notice that $|\bm{\omega}| \propto \gamma_0a$ (approximately)
for $\gamma_0a\gg1$.
This is the basis of the claim that the strengths 
of the spin resonances increase with the beam energy (to first order, anyway).
For the spin resonances driven by betatron oscillations,
the integrals symbolically reduce to expressions like
$\int \bm{\omega}_x\cdot\bm{k}_0 \propto \int x_\beta$ and
$\int \bm{\omega}_y\cdot\bm{k}_0 \propto \int y_\beta$.
(The integrands also have spin phase factors which I have omitted for now.)
As we progress to machines of higher energy,
the betatron tunes are $\gg1$ and increase with the design energy (as does the ring radius),
and the integrands oscillate rapidly and average efficiently to zero. 
Hence the spin resonance strength does {\em not} increase strongly with the design energy.
The same is true for $\int \bm{\omega}_{x,y}\cdot\bm{n}_0$, so higher order betatron spin resonances are weak.

The same is {\em not} true for the spin resonances driven by the synchrotron oscillations
(the term in $\bm{\omega}_\varepsilon$). 
Now $\int \bm{\omega}_\varepsilon\cdot\bm{k}_0 \propto \int \varepsilon$.
Even in very high energy rings such as FCC-ee, the synchrotron tune is still small (less than unity).
Although $Q_s = 0.65$ at $E=45.5$ GeV from the information in \cite{TLEPparams},
this is not yet large enough to claim `efficient averaging to zero.'
Hence for synchrotron resonances centered on an integer, we {\em can} say
$|\bm{\omega}| \propto \gamma_0a$ (approximately).
Similarly, $\int \bm{\omega}_\varepsilon\cdot\bm{n}_0 \propto \int \varepsilon$ is large, 
and the higher order synchrotron spin resonances are strong, as the data from LEP demonstrated.
Let us employ a smooth focusing approximation to make some estimates.
For simplicity, I treat only the term in the horizontal dispersion in $\bm{\omega}_\varepsilon$. 
Then, approximately, $G_x \simeq Q_x^2/R^2$ and $\eta_x \simeq R/Q_x^2$ so
\bq
\bm{\omega}_\varepsilon \propto (\gamma_0a+1)\, RG_x\eta_x
= (\gamma_0a+1) \,R\,\frac{Q_x^2}{R^2}\,\frac{R}{Q_x^2}
= \gamma_0a+1 \,.
\eq
As is well known, in an ideal planar ring 
$\bm{n}_0$ is vertical and $\bm{k}_0$ is horizontal, hence $\bm{e}_y\cdot\bm{k}_0=0$.
For the other term in $\bm{\omega}_\varepsilon$,
$\bm{e}_x\cdot\bm{k}_0\ne0$ but $\eta_y=1/\rho_y=0$ in an ideal planar ring.
The nonzero value of $\bm{\omega}\cdot\bm{k}_0$ arises from the lattice imperfections.
Hence, symbolically,
\bq
\zeta = -i \int \bm{\omega}_\varepsilon\cdot\bm{k}_0 \,d\theta^\prime
= (\gamma_0a+1) \times \textrm{(imperfection)} \times 
\frac{e^{i\,\textrm{(phase factor)}}}{\nu-\nu_{\rm res}}\times\varepsilon \,.
\eq
We can approximate the resonance denominator by $\frac12$ and ignore it below.
Then
\bq
\frac{\partial\zeta}{\partial\varepsilon} 
= (\gamma_0a+1) \times \textrm{(imperfection)} \times e^{i\,\textrm{(phase factor)}} \,.
\eq
Hence
\bq
\Bigl\langle\Bigl|\frac{\partial\zeta}{\partial\varepsilon}\Bigr|^2\Bigr\rangle
= (\gamma_0a+1)^2 \times \langle\textrm{(imperfection)}^2\rangle
\eq
This is the basis of the scaling law in eq.~\eqref{eq:polscale1},
for the dependence of the asymptotic polarization on the beam energy,
for first order spin resonances.
As the evidence 
from the LEP Energy Model demonstrated
(Fig.~3 in \cite{LEP_EPAC1994} and Fig.~10 in \cite{LEP_review})
this scaling law worked for a surprisingly large energy range,
across several rings, all fitted using the same value of $\alpha$ in eq.~\eqref{eq:polscale1}.

In Fig.~\ref{fig:pol_scaling},
I obtained an approximate fit to that data 
using $\alpha = 0.015$, for the data with harmonic spin matching
(see eq.~\eqref{eq:alpha_LEP}).
To extrapolate from LEP to FCC-ee, we need to analyze the imperfections.
For simplicity, I say there are approximately four times as many beamline elements in FCC-ee as in LEP,
because the circumference of FCC-ee (100 km) is about four times larger than LEP (26.7 km).
Then, assuming the imperfections in the individual beamline elements are uncorrelated,
I argue that 
\bq
\langle\textrm{(imperfection)}^2\rangle_{\rm FCC-ee} = 
\frac{\langle\textrm{(imperfection)}^2\rangle_{\rm LEP}}{4} \,.
\eq
Hence I claim that a reasonable value of $\alpha$ to use for FCC-ee,
in eqs.~\eqref{eq:polscale1} and \eqref{eq:polscale2},
is $\alpha = 0.015/\sqrt{4} = 0.015/2 \simeq 0.007$,
as stated in eq.~\eqref{eq:alpha_TLEP}.

\vfill\pagebreak
\subsection{Higher order spin resonances}
By 1992, I had realized that an algorithm to calculate the strengths of the higher order
spin resonances using a perturbation series in powers of the orbital amplitudes
was not satisfactory for high energy rings such as HERA and LEP.
I published an obscure paper in 1992 \cite{Mane_spinres_1992},
where I found that, 
using a program to calculate the spin integrals in a perturbation series in powers of the orbital amplitudes,
the spin resonances driven by the betatron oscillations were weak, 
even up to the energy of the $Z^0$ pole and beyond.
However, the synchrotron sideband resonances were too strong to be calculated perturbatively 
via a Taylor series in powers of the synchrotron oscillation amplitude.
I therefore suggested that a reasonable compromise would be to calculate the 
parent resonances perturbatively in powers of the orbital amplitudes, 
and then to incorporate the synchrotron sideband resonances
by multiplying each parent resonance by a depolarization enhancement factor.
This was clearly an approximate procedure, but should capture the essence of the spin resonance strengths.
I have never cited my 1992 paper in my subsequent work,
and I suspect it has no citations at all and played no role in the development of the LEP Energy Model.
Nevertheless, the data compiled by the team of the
LEP Energy Model suggests that the idea in that 1992 paper is essentially correct.

As I argued above, the most important set of spin resonances
are the synchrotron sideband spin resonances centered on an integer, say $n$.
Define $\Delta\nu = \nu_0 - n$ (where $\nu_0=\gamma_0a$ in a planar ring).
If $\bm{\omega}\cdot\bm{k}_0$ has a Fourier harmonic with a coefficient $a_n$,
then Yokoya \cite{Yokoya_syncsdb_1983}
derived the following expression for the 
synchrotron sideband spin resonances centered on an integer
(eq.~(3.17) in \cite{Yokoya_syncsdb_1983})
\bq
\Bigl\langle\Bigl|\frac{\partial\zeta}{\partial\varepsilon}\Bigr|^2\Bigr\rangle
= |a_n|^2 \,\sum_{m=-\infty}^\infty 
\frac{(\Delta\nu)^2}{[(\Delta\nu +mQ_s)^2 - Q_s^2]^2}\,e^{-\sigma^2}\,I_m(\sigma^2) \,.
\eq
Here $I_m$ is a modified Bessel function.
The synchrotron tune modulation index $\sigma$ is given by
\bq
\sigma = \frac{\partial\nu}{\partial\varepsilon}\,\frac{\sigma_\varepsilon}{Q_s}
= \frac{\gamma_0 a\,\sigma_\varepsilon}{Q_s} \,.
\eq
The last expression is for a planar ring, where $\nu = \gamma a$.
I tabulate the values of the tune modulation index in Table \ref{tab:ring_params}.
The `depolarization enhancement factor' $\mathcal{F}$ 
is given by dividing the above by the strength of the parent resonance
(obtained by setting $\sigma=0$), i.e.
\bq
\label{eq:depolenfac}
\mathcal{F} = 
\Bigl\langle\Bigl|\frac{\partial\zeta}{\partial\varepsilon}\Bigr|^2\Bigr\rangle
\Bigl/
\Bigl\langle\Bigl|\frac{\partial\zeta}{\partial\varepsilon}\Bigr|^2\Bigr\rangle_{\sigma=0}
= [(\Delta\nu)^2 - Q_s^2]^2 \,\sum_{m=-\infty}^\infty 
\frac{e^{-\sigma^2}\,I_m(\sigma^2)}{[(\Delta\nu +mQ_s)^2 - Q_s^2]^2} \,.
\eq
This depends only on known or easily calculated parameters, not on imperfections, 
hence it can be accurately computed for FCC-ee.
We can fix $\Delta\nu = 0.5$ in numerical estimates for FCC-ee (working point halfway between two integers).

I employed the above expression for $\mathcal{F}$ in eq.~\eqref{eq:polscale2},
to plot the graphs in Fig.~\ref{fig:pol_scaling}.
For simplicity I fixed the synchrotron tune at $Q_s = 0.065$ at all energies.
I also fixed $C = 26.7$ km and $\rho = 3.1$ km, i.e.~the values for LEP.

To compute $\mathcal{F}$ for FCC-ee, 
it is not adequate to include only a single parent resonance.
For example, from \cite{TLEPparams} (see also Table \ref{tab:ring_params}), 
the synchrotron tune is $Q_s=0.65$ in FCC-ee at the $Z^0$ pole.
Hence at any given working point, sideband resonances from
several parent resonances (several values of $n$) will contribute.
I calculated $\mathcal{F}$ by summing over several integers $n$,
where I assumed $|a_n|^2 = |a_{n\pm1}|^2 = |a_{n\pm2}|^2$, etc.
This should be a reasonable approximation at high energies and large values of $n$.
Hence I used the following expression in my numerical estimates for FCC-ee
\bq
\label{eq:depolenfac1}
\mathcal{F} = \frac{[(\Delta\nu)^2 - Q_s^2]^2}{(\Delta\nu)^2} \,\sum_{m=-50}^{50} \sum_{k=-3}^3
\frac{(k+\Delta\nu)^2}{[(k+\Delta\nu +mQ_s)^2 - Q_s^2]^2}\,e^{-\sigma^2}\,I_m(\sigma^2) \,.
\eq
I found that using more values of $m$ and $k$ did not affect the numerical results significantly.

\vfill\pagebreak
\section{Estimates for polarization in FCC-ee}
\subsection{Formula}
Hence I suggest the following overall approximate expression to estimate the polarization in FCC-ee,
including both the wigglers and the spin resonances
\begin{align}
\label{eq:polscale_all}
\frac{P}{P_{ST}} 
&= \frac{A_n}{A_p(1 + (\alpha E)^2\mathcal{F})} 
= \frac{1 + 0.006\,\mathcal{N}^3}{(1 + 0.01\,\mathcal{N}^3)(1 + (\alpha E)^2\mathcal{F})} \,,
\\
\label{eq:tauscale_all}
\tau_p 
&= \frac{\tau_{ST}}{A_p(1 + (\alpha E)^2\mathcal{F})} 
= \frac{\tau_{ST}}{(1 + 0.01\,\mathcal{N}^3)(1 + (\alpha E)^2\mathcal{F})} \,.
\end{align}
\bit
\item
This model arbitrarily assumes $10^4$ arc dipoles and $100$ wigglers.
The arcs are isomagnetic.
All the wigglers have the same design; it is an `isowiggler' model.
\item
The weighted average using the luminosity lifetime also needs to be taken into account.
\item
For the parent spin resonances,
I suggest that $\alpha \simeq 0.007$ is a reasonable estimate for FCC-ee.
The model includes only synchrotron sideband spin resonances centered on an integer.
The quantitative expression for the depolarization enhancement factor $\mathcal{F}$
is given in eq.~\eqref{eq:depolenfac1}.
Note that the value of $\mathcal{F}$ depends on the synchrotron tune
and the energy spread.
The synchrotron tune varies with the beam energy (see \cite{TLEPparams}).
The energy spread depends on the beam energy
and the wigglers increase the energy spread.
Hence the value of $\mathcal{F}$ depends on the value of $\mathcal{N}$.
\item
Conversely, the value of $\mathcal{N}$ depends on the value of $\mathcal{F}$.
At any given beam energy, the speedup factor for the polarization time 
(using the wigglers) depends on the value of $\mathcal{N}$.
However, varying the value of $\mathcal{N}$ changes the energy spread, 
thence the value of the depolarization enhancement factor $\mathcal{F}$,
which then changes the polarization time.
Hence the appropriate value of $\mathcal{N}$ must be calculated self-consistently.
\item
In addition to all of the above,
there may be constraints on the radiated power or tolerable energy spread.
\eit

\subsection{$Z^0$ pole}
Let us examine what happens at $E = 45.5$ GeV.
From Table \ref{tab:ring_params}, the polarization time is about 270 h, without wigglers and spin resonances.
First consider the situation with the spin resonances but without wigglers.
From Table \ref{tab:ring_params}, the tune modulation index is $\sigma \simeq 0.095$.
This is a small value.
Also $\alpha E \simeq 0.3185$ and $\mathcal{F} \simeq 1.11$,
where I set $\Delta\nu=0.5$.
Then the asymptotic polarization and the polarization time are
\bq
P \simeq 83.0 \% \,,\qquad
\tau_p \simeq 243 \ \textrm{h} \,.
\eq
(For the record, the luminosity limetime is 213 min
(see \cite{TLEPparams} and Table \ref{tab:ring_params}) 
and the lifetime weighted polarization average is about $1.2\%$.)
Hence we seek a speedup factor of 243 (as opposed to 270), to reduce the polarization time to about 1 h.
Let us set $\mathcal{N} = 28$. 
The numerical estimates for the asymptotic polarization and the time constant, 
etc.~are given in the first row of Table \ref{tab:pol_estimates}.
The estimated asymptotic polarization is about 50\%,
which is comparable to the best that was attained at LEP at the $Z^0$ pole (57\%).
The lifetime weighted polarization average is about $38.0\%$.
The radiated energy per turn increases by a factor of about 9, 
so the single beam current must decrease by this factor to maintain the total radiated power at 100 MW.
The relative energy spread increases to $0.19\%$.

\subsection{$W$ pair threshold}
Next, let us examine what happens at $E = 80$ GeV.
From Table \ref{tab:ring_params}, the polarization time is about 16 h, without wigglers and spin resonances.
First consider the situation with the spin resonances but without wigglers.
From Table \ref{tab:ring_params}, the tune modulation index is $\sigma \simeq 0.78$.
This is a large value.
Also $\alpha E \simeq 0.56$ and $\mathcal{F} \simeq 6.28$.
Then the asymptotic polarization and the polarization time are
\bq
P \simeq 31.1 \% \,,\qquad
\tau_p \simeq 5.4 \ \textrm{h} \simeq 325 \ \textrm{min} \,.
\eq
Hence we seek a speedup factor of $5.5$, which is a modest value, to reduce the polarization time to 1h.
Let us set $\mathcal{N} = 7$. 
The numerical estimates for the asymptotic polarization and the time constant, 
etc.~are given in the second row of Table \ref{tab:pol_estimates}.
The estimated asymptotic polarization is about 17\%,
which is comparable to what was attained at LEP at the $Z^0$ pole with harmonic spin matching.
The luminosity lifetime is about 52 min and the polarization time is about 57 min,
so the lifetime weighted polarization average is about $7.9\%$.
Unfortunately this is marginal for resonant depolarization measurements.
Remember there are additional sources of depolarization I have not taken into account.

\subsection{Non-colliding pilot bunches}
At the $Z^0$ pole, it seems that one can attain adequate polarization using colliding bunches,
but the polarization at the $W$ pair production threshold is marginal for resonant depolarization measurements.
Hence let us consider the use of non-colliding pilot bunches.
They will not be topped up and will have a longer beam lifetime.
I reanalyze what happens at the $W$ pair production threshold.
I shall suppose the beam lifetime of the non-colliding bunches is infinite.
We saw above that with $\mathcal{N} = 7$,
the estimated asymptotic polarization was about 17\% and the polarization time was about 57 min.
Let us employ a more relaxed value of $\mathcal{N} = 5$ for the wigglers.
The numerical estimates for the asymptotic polarization and the time constant, 
etc.~are given in the third row of Table \ref{tab:pol_estimates}.
The estimated asymptotic polarization is about 21\% and the polarization time is about 127 min.
These results may be acceptable for practical use.
The radiated energy per turn increases by a factor of about 1.25, 
which is lower than the value of $1.5$ for $\mathcal{N}=7$.
The relative energy spread increases to $0.11\%$, as compared to $0.13\%$ for $\mathcal{N}=7$.
Hence the use of non-colliding bunches may be helpful at the $W$ pair production threshold.

\subsection{Higgs threshold}
Next, let us examine what happens at $E = 120$ GeV.
From Table \ref{tab:ring_params}, the polarization time is about $2.1$ h, without wigglers and spin resonances.
First consider the situation with the spin resonances but without wigglers.
From Table \ref{tab:ring_params}, the tune modulation index is $\sigma \simeq 3.98$.
This is a very large value.
Also $\alpha E \simeq 0.84$ and $\mathcal{F} \simeq 158.2$.
Note from \cite{TLEPparams} (see also Table \ref{tab:ring_params}), the synchrotron tune is $Q_s=0.096$.
Hence setting $\Delta\nu = 0.5$ would put the working point close to the center of a spin resonance.
Hence I set $\Delta\nu = 0.45$ to calculate $\mathcal{F}$.
Then the asymptotic polarization and the polarization time are
\bq
P \simeq 0.82 \% \,,\qquad
\tau_p \simeq 0.02 \ \textrm{h} \simeq 1.1 \ \textrm{min}\,.
\eq
The level of the asymptotic polarization is too low to be useful
and using wigglers will only make things worse.
The polarization time is about 1 min.
This will be so even for non-colliding bunches.
Hence there is no useful polarization at this energy, and the same will be true at even higher energies.

\subsection{Summary}
At the $Z^0$ pole, there should be no problem to obtain a useful degree of polarization, 
with a reasonable time constant, even with colliding bunches.
The situation is not so clear at the $W$ pair production threshold.
The use of non-colliding pilot bunches may be required at a beam energy of 80 GeV.
At higher energies, such as the threshold of Higgs production ($E = 120$ GeV),
the spin resonances will likely reduce the asymptotic polarization to a level that is not useful.  
For energies where a useful degree of polarization is attainable,
a separate analysis of machine operations will be required,
to determine if the reduction of luminosity (single beam current),
and the increase in the energy spread, etc.~are acceptable.

\vfill\pagebreak
\section{Alternative Ideas}
\subsection{General remarks}
It may be possible to obtain a useful degree of polarization,
with a reasonable time constant, at very high energies ($E \ge 80$ GeV).
The root cause of the difficulty in obtaining
useful polarization in very high energy $e^+e^-$ rings
is the large value of the spin tune modulation index.
This makes the spin resonances too strong.
Recall the definition spin tune modulation index is
\bq
\sigma = \frac{\partial\nu}{\partial\varepsilon}\,\frac{\sigma_\varepsilon}{Q_s} \,.
\eq
If one thinks about it, $(\partial\nu/\partial\varepsilon)\,\sigma_\varepsilon$
is just the r.m.s.~spin tune spread,
and the spin tune modulation index is simply
\bq
\sigma = \frac{\textrm{r.m.s.~spin tune spread}}{\textrm{spacing between resonances}} \,.
\eq
In other words, it is the ratio of the r.m.s.~spin tune spread
to the spacing between the synchrotron sideband resonances,
which are the strongest of the spin resonances.
In a planar ring, the spin tune is $\nu = \gamma a = \gamma_0a(1+\varepsilon)$ and
\bq
\frac{\partial\nu}{\partial\varepsilon} = \gamma_0 a \,.
\eq
This leads to the value in a planar ring
\bq
\sigma_{\rm planar} = \frac{\gamma_0 a\,\sigma_\varepsilon}{Q_s} \,.
\eq
This is a large value at very high energies, 
and that makes the synchrotron sideband spin resonances very strong, 
and that is the source of the difficulties.
  
Most of the difficulties to obtain useful polarization in very high energy
$e^+e^-$ rings would be solved, or at least greatly alleviated,
if the magnitude of the spin tune modulation index could be reduced.
As I found from my studies in 1992  \cite{Mane_spinres_1992}, 
the other spin resonances (those driven by the betatron oscillations) 
are relatively much weaker and should not be problematical in FCC-ee.

\subsection{Basic scheme}
I suggest the following scheme, using a combination of Siberian Snakes and wigglers.
Siberian Snakes reduce the magnitude of the spin tune modulation index, 
but Snakes come with side-effects, which I shall discuss below.
The proposed scheme is to employ two Snakes,
which partition the bending in the circumference into fractions $f$ and $1-f$.
Hence the Snakes are {\em not} at diametrically opposite points in the ring.
The Snakes have {\em parallel} spin rotation axes.
The Sokolov-Ternov polarization will be decreased, 
by a factor $1-2f$ (in an isomagnetic ring),
because the spins will point up on one side and down on the other side.
To remedy this, I employ wigglers with {\em opposite} polarity on the two sides,
i.e.~polarity $-1:\frac12:-1$ in one arc and the opposite $+1:-\frac12:+1$ in the other arc.
Then $\hat{\bm{b}}\cdot\bm{n}_0$ will have the {\em same} sign in all the wigglers all around the ring.
Hence the polarization will be generated by the wigglers.
There are several issues to be addressed in this scheme:
\bit
\item
Depolarization and increase of the vertical emittance caused by the Snakes themselves.
The Snakes will contain horizontal (radial) magnetic fields.
\item
The effect of the Snakes on the spin tune.
Theoretical Snakes are ideal $180^\circ$ spin rotators,
but real Snakes will yield a systematic error in the spin tune.
\item
The contribution of other sources of error such as the Earth and lunar tides,
hydrogeologic effects, etc.~for resonant depolarization.
\item
The reduction of the spin tune modulation index.
\item
The asymptotic polarizatoin and polarization time, 
in particular at the highest energies.
\item
As always with wigglers, the increase in the radiated power and the beam energy spread.
\eit

\vfill\pagebreak
\subsection{Snake designs}
Is a practical design of Snakes possible at FCC-ee?
I shall argue that the answer is {\em yes}.
The simplest design of a Siberian Snake is a solenoid, which rotates the spin by $180^\circ$ around the beam axis.
The transverse $x-y$ coupling will be compensated by skew quadrupoles.
The required integrated magnetic is given by
\bq
(1+a)\,\frac{BL}{B\rho} = \pi\,.
\eq
Hence $BL \propto B\rho \propto E$, which makes solenoids impractical at high energies. 
The required magnetic field increases with energy, and is too high to be practical at FCC-ee.
The alternative is to employ combinations of transverse magnetic fields.
Then we obtain a relation of the form
\bq
(\gamma_0a+1)\,\frac{BL}{B\rho} = \textrm{spin rotation angle} \,.
\eq
Hence, for $\gamma_0a\gg1$,
\bq
BL \propto \textrm{const}\,\frac{B\rho}{\gamma_0a+1} \propto \frac{p}{E} = \frac{v}{c} \,.
\eq
The required magnetic field scales in proportion to $v/c$.
This is precisely what happens for the Snakes at RHIC.
In practice, the variation of $v/c$ is so small at RHIC, from injection to flattop,
that the Snake magnetic fields are held a fixed value at all energies.
The same will be the case for FCC-ee; the same value of $BL$ will work at all operating energies.
It is well known that the transverse orbit deflections scale as $1/E$ for $\gamma_0a\gg1$
\bq
\frac{BL}{B\rho} \propto \frac{\textrm{const}}{\gamma_0a+1} \propto \frac{1}{E} \,.
\eq
This is precisely what happens in the RHIC Snakes;
the orbit excursions are largest at injection and decrease at higher energies.
Let us analyze the radiation loss per turn in a Snake with transverse magnetic fields.
The radiation loss over a length $L$ is given by
\bq
U = \textrm{const} \times E^2 \langle B^2\rangle LI \,.
\eq
Here $E$ is the beam energy, 
$\langle B^2\rangle$ is the mean-square magnetic field, 
$L$ is the length and $I$ is the beam current.
Hence for a fixed Snake magnetic field $B_{\rm snake}$ and length $L_{\rm snake}$,
\bq
U_{\rm snake} \propto E^2 \,.
\eq
Compare this to the radiation loss per turn in the arcs (isomagnetic ring)
\bq
U_0 = \textrm{const}\times\frac{E^4}{\rho} \propto E^4 \,.
\eq
Hence the radiation in a Snake increases only as $E^2$ and the {\em relative} radiation loss 
(relative to the arcs) scales as $1/E^2$.
Hence if a Snake design will work at $E=80$ GeV, say, then it will work even better at higher energies.

One design of Snakes is to employ combinations of horizontal and vertical bends.
For example Steffen Snakes have a structure of interleaved H and V bends
\bc
$\cdots$ --H\ --V\ 2H\ 2V\ --2H\ -V\ H\ $\cdots$
\ec
The presence of vertical bends causes the synchrotron radiation to increase the vertical emittance.
The radiation can also drive spin resonances, because of the nonzero horizontal and vertical dispersion in the Snake.
The condition to be a Snake is $\sin^2\psi_H \sin^2\psi_V = \frac12$,
so a Steffen Snake can be realized by setting 
$\psi_H = \psi_V = \sin^{-1}(2^{-1/4}) \simeq 0.9989 \simeq 1$.
Then the orbit rotation angle, at a beam energy of $80$ GeV, is 
\bq
\theta_b = \frac{BL}{B\rho} \simeq \frac{1}{80/0.440} \simeq 0.0055 = 5.5 \;\textrm{mrad} \,.
\eq
If each magnet is 10 m long, the vertical orbit excursion in the central Vbend is about $50 \times 0.0055 \simeq 27.5$ cm.
In practice, a dipole of length 10 m and bend angle of 4 mrad will have a bend radius of 2500 m, 
which will generate a lot of radiation.
If instead each magnet is 100 m long (say ten dipoles in series), 
the vertical orbit excursion in the central Vbend will increase to about $27.5$ m.
These are inconvenient numbers.
Orbit excursions of this magnitude cannot fit in a straight beam pipe.

A better design of Snakes is to employ helical magnetic fields.
This is how the RHIC Snakes are constructed.
To the leading order, the helical magnetic field is
\bq
B_y = B_h\,\cos(ks) \,,\qquad
B_x = \eta_h\,B_h\,\sin(ks) \,.
\eq
Let the `wavelength' of the helical field be $\lambda_h$, then the helix pitch is $k=2\pi/\lambda_h$.
If $\eta_h=\pm 1$ the twist of the helix has positive/negative helicity.
I shall employ the Snake design of the BINP team (Budker Institute of Nuclear Physics)
\cite{Ptitsin_Shatunov_HelicalSnakes_NIMA}.
It was employed for the RHIC Snakes \cite{HelicalSnakes_PAC95}.
The design consists of four equal length helical modules, all with helicity $\eta=1$
with magnetic fields in the pattern $(B_1,B_2,-B_2,-B_1)$.
This field pattern automatically yields a spin rotation axis in the horizontal plane.
In each module, the total twist angle of the helix is $720^\circ$ (in the RHIC Snakes it is $360^\circ$).
Hence each helix module contains two full twists.
There are important reasons why the twist angle should be a multiple of $360^\circ$,
see \cite{Ptitsin_Shatunov_HelicalSnakes_NIMA} or the review \cite{MSY1} for details.
In particular, it guarantees automatic internal closure of the orbit excursions in the helix;
no compensating dipoles or quads, etc.~are required.
Furthermore, optically the helices look like drift spaces of the same physical length that they occupy.
Hence, for example, RHIC can be operated with and without the Snakes and spin rotators, 
without change to the rest of the machine optics.
The radius of the helical orbit in the Snake is given by
\bq
r_h = \frac{1}{k^2}\,\frac{eB_h}{B\rho} \,.
\eq
The ratio of the radiated energy is
\bq
\frac{U_h}{U_{\rm arc}} = \frac{L_h B_h^2}{L_{\rm arc}B_{\rm arc}^2}
= \frac{L_h B_h^2}{2\pi\rho_0\,(B\rho)^2/\rho_0^2}
= \frac{\rho_0 L_h B_h^2}{2\pi(B\rho)^2} \,.
\eq
I employed a design with $\lambda=100$ m, so each module length is 200 m, and the overall Snake length is 800 m.
I set $B_2=-B_1$. 
The overall spin rotation angle depends on the value of $B_1$.
I set the beam energy to 80 GeV.
I obtained a Snake, i.e.~$180^\circ$ spin rotation angle, using $B_1 \simeq 0.034$ T, i.e.~340 Gauss.
The spin rotation axis points about $8.8^\circ$ from the beam axis.
The maximum orbit excursion is about $3.3$ cm, which should fit in a straight beam pipe.
The (ratio of the) radiation emitted in the Snake is
\bq
\frac{U_{\rm snake}}{U_{\rm arc}} \simeq 0.0228 \simeq 2.3 \% \,.
\eq
Other parameter values are of course possible.
The main point is to argue
that a feasible set of helical Snake parameters {\em is} possible at FCC-ee.
Note that we need two Snakes, and each Snake is optically a drift space of 800 m.
The ring lattice must be designed to accomodate this.
Nevertheless, the required magnetic field is modest, the orbit excursion is small 
and the vertical dispersion (hence any excitation of the vertical emitttance and spin resonances) is likely to be small.
As I have pointed out above, the scalings of the orbit excursion, etc.~with energy
will make the performance even better at higher beam energies.

\vfill\pagebreak
\subsection{Spin tune}
For a pair of Snakes with spin rotation axes oriented at angles $\xi_1$ and $\xi_2$
relative to the beam axis, the spin tune (for ideal Snakes) is
\bq
\nu = \frac{\xi_2-\xi_1}{\pi} +(1-2f)\gamma a \,.
\eq
If $\xi_2-\xi_1 = \frac12\pi$ (orthogonal Snakes axes) 
and $f=\frac12$ (diametrically opposed Snakes),
the spin tune is $\nu=\frac12$, as is well known.
However, I choose parallel Snake axes $\xi_1=\xi_2$, so the constant term vanishes.
Then
\bq
\nu = (1-2f)\gamma a + \textrm{(systematic error from Snakes)} \,.
\eq
I shall discuss the systematic error later.
The use of parallel Snake axes means the two Snakes can be built to the same design,
and be powered in series, which will help to constrain the systematic error.
I suggest setting $f = 0.48$. Then $1-2f = 0.04$.
The spin tune will be proportional to the beam energy, and can be used for energy calibration.

\subsection{Resonant depolarization}
Note that if $1-2f = 0.04$, 
the effects of systematic errors such as lunar and Earth tides,
water levels and hydrogeology, etc.~will be magnified by a factor of $1/0.04 = 25$.
This must be taken into account in the practical measurements of resonant depolarization (RDP)
because
\bq
E = \frac{m_ec^2}{a}\,\frac{\nu_{\rm RDP} - \Delta\nu_{\rm syst}}{1-2f} 
= \frac{m_ec^2}{a}\,\frac{f_{\rm RDP}-\Delta f_{\rm syst}}{f_{\rm rev}}\,\frac{1}{1-2f} \,.
\eq
The effects of all systematic and statistical errors in the measurement of $f_{\rm RDP}$
will be magnified by a factor of $1/(1-2f)$.

I found that $f=0.48$ was a reasonable compromise value.
Using a larger value such as $f=0.49$ did not increase the asymptotic polarization significantly,
and magnified the sensitivity of the energy calibration to systematic and statistical errors.
Using a smaller value such as $f=0.45$ decreased the asymptotic polarization noticeably at the highest energies.
Hence I suggest that $f=0.48$ is a good value to employ.

\vfill\pagebreak
\subsection{Systematic error from Snakes}
I perform a simple analysis of the systematic error in the spin tune induced by the Snakes.
Since the Snakes have parallel spin rotation axes,
I shall treat them as identical.
Suppose the Snake spin rotation axis points along the beam axis.
Let the spin rotation angle of the Snakes be $\pi(1 +\delta)$, where $|\delta|\ll1$.
Let the coordinate axes be $(\bm{e}_1,\bm{e}_2,\bm{e}_3)$, which are radial, longitudinal and vertical, respectively.
Technically, I need the Snake axes to be antiparallel, 
so the Snake spin rotation matrices are
$M_{\rm S1} = e^{-i\pi(1+\delta)\sigma_2/2}$ and $M_{\rm S2} = e^{i\pi(1+\delta)\sigma_2/2}$.
The one turn spin rotation matrix, on the design orbit, is
\bq
\begin{split}
M &= e^{i\pi(1+\delta)\sigma_2/2}
\,e^{-i\pi\nu_0 f\sigma_3}
\,e^{-i\pi(1+\delta)\sigma_2/2}
\,e^{-i\pi\nu_0(1-f)\sigma_3}
\\
&= \Bigl(\sin\frac{\pi\delta}{2} -i\,\cos\frac{\pi\delta}{2}\,\sigma_2 \Bigr)
\,e^{-i\pi\nu_0 f\sigma_3}
\Bigl( \sin\frac{\pi\delta}{2} +i\,\cos\frac{\pi\delta}{2}\,\sigma_2 \Bigr)
\,e^{-i\pi\nu_0(1-f)\sigma_3}
\\
&=
\sin^2\frac{\pi\delta}{2} \,e^{-i\pi\nu_0\sigma_3}
+\cos^2\frac{\pi\delta}{2} \,e^{-i\pi\nu_0(1-2f)\sigma_3}
\\
&\quad
+i\,\sin\frac{\pi\delta}{2}\cos\frac{\pi\delta}{2} \,\sigma_2\,
\Bigl[\,e^{-i\pi\nu_0(1-2f)\sigma_3} -e^{-i\pi\nu_0\sigma_3} \,\Bigr] \,.
\end{split}
\eq
The spin tune is obtained from half the trace
\bq
\cos(\pi\nu) = 
\sin^2\frac{\pi\delta}{2} \,\cos(\pi\nu_0)
+\cos^2\frac{\pi\delta}{2} \,\cos(\pi\nu_0(1-2f)) \,.
\eq
For $\delta = 0$ we see that $\nu = \nu_0(1-f) = \gamma_0a(1-2f)$.
Next
\bq
\frac{\partial\nu}{\partial\delta} = 
-\frac{\sin(\pi\delta)}{2} \,\frac{\cos(\pi\nu_0) -\cos(\pi\nu_0(1-2f))}{\sin(\pi\nu)}
= O(\delta) \,.
\eq
Hence for small $|\delta|$
\bq
\begin{split}
\nu &\simeq \gamma_0a(1-2f) 
-\frac{1-\cos(\pi\delta)}{2\pi} \,\frac{\cos(\pi\nu_0) -\cos(\pi\nu_0(1-2f))}{\sin(\pi\nu_0(1-2f))}
\\
&= \gamma_0a(1-2f) + O(\delta^2) \,.
\end{split}
\eq
The systematic error arising from $|\delta| \ll 1$ is of second order $O(\delta^2)$.
Hence it should be a very small systematic error.
It should be possible to configure the Snakes such that $|\delta|\le 10^{-3}$,
for example by varying the Snake currents and making a map of the spin tune.
In practice there are also other sources of systematic error.
For example, the Snake axes will not, in general, point exactly along the beam axis.
Also, the Snake axes will not be exactly (anti)parallel.
All of these effects need to be analyzed, but they should all be small.

\vfill\pagebreak
\subsection{Spin tune modulation index and polarization}
Neglecting the systematic error from the Snakes, we see that
\bq
\frac{\partial\nu}{\partial\varepsilon} = (1-2f)\gamma_0 a \,.
\eq
The systematic error will also depend on $\varepsilon$,
but it should be very small, much less than $(1-2f)\gamma_0 a$.
Compared to the situation in a planar ring, the spin tune modulation index will be
\bq
\sigma_{\rm Snakes} = (1-2f)\,\frac{\gamma_0 a\,\sigma_\varepsilon}{Q_s} 
= (1-2f)\,\sigma_{\rm planar} \,.
\eq
If we set $f = 0.48$, 
the value of the spin tune modulation index will be reduced by a factor of 25.
The smaller value of $\sigma$ will reduce the value of the 
depolarization enhancement factor $\mathcal{F}$,
i.e.~weaken the synchrotron sideband spin ressonances,
which is the fundamental idea underlying this scheme.

For the asymptotic polarization, in an isomagnetic ring the 
Sokolov-Ternov polarization level will be multiplied by a factor $1-2f$.
Hence for $f=0.48$, the asymptotic polarization will not be more than 
$0.04 \times 8/(5\sqrt3) \simeq 3.7\%$.
The wigglers will be required to increase the asymptotic polarization,
as well as to control the polarization time.
I define a new numerator term for the polarization formula (`$s$' for `Snakes')
\bq
A_s = 1-2f + 0.006\,\mathcal{N}^3 \,.
\eq
The overall polarization formula is 
\begin{align}
\label{eq:polscale_snak}
\frac{P}{P_{ST}} 
&= \frac{A_s}{A_p(1 + (\alpha E)^2\mathcal{F})} 
= \frac{1-2f + 0.006\,\mathcal{N}^3}{(1 + 0.01\,\mathcal{N}^3)(1 + (\alpha E)^2\mathcal{F})} \,,
\\
\label{eq:tauscale_snak}
\tau_p 
&= \frac{\tau_{ST}}{A_p(1 + (\alpha E)^2\mathcal{F})} 
= \frac{\tau_{ST}}{(1 + 0.01\,\mathcal{N}^3)(1 + (\alpha E)^2\mathcal{F})} \,.
\end{align}
The formula for the polarization time is not changed from that in eq.~\eqref{eq:tauscale_all},
but the value of $\mathcal{F}$ will be smaller.

\vfill\pagebreak
\subsection{Estimated polarization}
\noindent{\bf $W$ pair threshold}\\
I analyse what happens at $E = 80$ GeV.
This is the scenario where the orbit excursions and the relative radiation in the Snakes will have their largest values.
From Table \ref{tab:ring_params}, the polarization time is about $16.1$ h, without wigglers and spin resonances.
First consider the situation with the spin resonances but without wigglers.
The tune modulation index is $\sigma \simeq 0.031$, very small.
From \cite{TLEPparams} (see also Table \ref{tab:ring_params}), the synchrotron tune is $Q_s=0.21$.
Hence I set $\Delta\nu = 0.5$.
Also $\alpha E \simeq 0.56$ and $\mathcal{F} \simeq 1.13$, also small.
Then the asymptotic polarization and the polarization time are
\bq
P \simeq 2.7 \% \,,\qquad
\tau_p \simeq 11.9 \ \textrm{h} \,.
\eq
We seek a speedup factor of about 12, to reduce the polarization time to 1h.
Let us set $\mathcal{N} = 10$. 
The numerical estimates for the asymptotic polarization and the time constant, 
etc.~are given in the fifth row of Table \ref{tab:pol_estimates}.
The results are encouraging.
The asymptotic polarization is about 38\% with a time constant of about 39 min.
From \cite{TLEPparams} (see also Table \ref{tab:ring_params}), the luminosity lifetime is about 52 min,
hence the lifetime weighted average polarization is about $22\%$.
This is acceptable for energy calibration measurements.
It is of approximately the same quality as 
what I estimated above for non-colliding bunches at the $W$ pair production threshold,
which was an asymptotic polarization of about 21\% with a time constant of about 2 h.
Admittedly, the radiation loss per turn is larger, so the beam current (hence luminosity) must be reduced.
However, with Snakes, one can achieve this performance even for colliding bunches.
The relative energy spread is about $0.19\%$,
about a factor of 2 larger than the value of $0.09\%$ from \cite{TLEPparams}.
(Both of the above figures include the beamstrahlung contribution.)
The radiation loss per turn increases by a factor of $2.44$,
hence the single beam current must be decreased by the same factor,
to maintain the radiated power at 100 MW.
Hence the radiation from each Snake will be about $0.0228 \times 50/2.44 \simeq 0.47$ MW over a length of 800 m.
This should be acceptable,
because the nominal FCC-ee design calls for 50 MW per ring over 100 km, i.e.~$0.5$ MW per km.
Recall from above that the maximum orbot excursion is about 3.3 cm.

\noindent{\bf Higgs threshold}\\
Next let us examine what happens at the Higgs production threshold, $E = 120$ GeV.
From Table \ref{tab:ring_params}, the polarization time is about $2.1$ h, without wigglers and spin resonances.
First consider the situation with the spin resonances but without wigglers.
The tune modulation index is $\sigma \simeq 0.16$, a small value.
From \cite{TLEPparams} (see also Table \ref{tab:ring_params}), the synchrotron tune is $Q_s=0.096$.
Hence to avoid a working point close to the center of a spin resonance I set $\Delta\nu = 0.45$.
Also $\alpha E \simeq 0.84$ and $\mathcal{F} \simeq 1.16$, still acceptable.
Then the asymptotic polarization and the polarization time are
\bq
P \simeq 2.0 \% \,,\qquad
\tau_p \simeq 1.17 \ \textrm{h} \simeq 70 \ \textrm{min}\,.
\eq
From \cite{TLEPparams} (see also Table \ref{tab:ring_params}), 
the luminosity lifetime is about 21 min.
Hence we seek to speed up the polarization time by a factor of 3.
Let us set $\mathcal{N} = 7$. 
The numerical estimates for the asymptotic polarization and the time constant, 
etc.~are given in the sixth row of Table \ref{tab:pol_estimates}.
The results are encouraging.
The time constant is about 16 min, about the same as the luminosity lifetime.
The asymptotic polarization is about 24\% 
and the lifetime weighted average polarization is about 14\%.
This should be acceptable for resonant depolarization measurements.
The relative energy spread is about $0.19\%$,
which is only slightly larger than the value from
\cite{TLEPparams}, which is $0.14\%$.
Again, both of the above figures include the beamstrahlung contribution.
The radiation loss per turn again increases by a modest factor of $1.25$,
hence the single beam current must be decreased by the same factor,
to maintain the radiated power at 100 MW.

The maximum orbit excursion in the Snakes is about $2.2$ cm
and the relative radiation loss is about $1.0\%$.
Hence the radiation from each Snake will be about $0.01 \times 50/1.49 \simeq 0.32$ MW over a length of 800 m.
These values should be acceptable.

\vskip 0.2in
\noindent{\bf $t\bar{t}$ threshold}\\
Next let us examine what happens at the highest energy of interest,
viz.~the $t\bar{t}$ threshold $E = 175$ GeV.
From Table \ref{tab:ring_params}, the polarization time is about 19 min, without wigglers and spin resonances.
First consider the situation with the spin resonances but without wigglers.
The tune modulation index is $\sigma \simeq 0.302$, which is tolerable
(about the same as in SPEAR at $3.5$ GeV).
From \cite{TLEPparams} (see also Table \ref{tab:ring_params}), the synchrotron tune is $Q_s=0.1$.
Hence setting $\Delta\nu = 0.5$ would put the working point at the center of a spin resonance.
Hence to calculate $\mathcal{F}$, I set $\Delta\nu = 0.45$, to lie halfway between two synchrotron sideband resonances.
Also $\alpha E \simeq 1.225$ and $\mathcal{F} \simeq 1.2$, which is small.
Then the asymptotic polarization and the polarization time are
\bq
P \simeq 1.3 \% \,,\qquad
\tau_p \simeq 0.14 \ \textrm{h} \simeq 6.8 \ \textrm{min} \,.
\eq
The polarization time is already about 7 min.
We therefore seek a modest setting for the wigglers.
Let us set $\mathcal{N} = 5$. 
The numerical estimates for the asymptotic polarization and the time constant, 
etc.~are given in the seventh row of Table \ref{tab:pol_estimates}.
The time constant is about 3 min while the luminosity lifetime is about 15 min.
The asymptotic polarization is about $11.3\%$
and the lifetime weighted average polarization is about $9.5\%$, 
which is almost the same because the luminosity lifetime is much longer than the polarization time.
The above numbers should be accepatble for resonant depolarization measurements.
The relative energy spread is about $0.23\%$,
which is only slightly larger than the value from
\cite{TLEPparams}, which is $0.19\%$.
Note that both of the above figures include the beamstrahlung contribution.
The radiation loss per turn increases by a modest factor of $1.25$,
hence the single beam current must be decreased by the same factor,
to maintain the radiated power at 100 MW.

The maximum orbit excursion in the Snakes is about $1.5$ cm
and the relative radiation loss is about $0.5\%$.
Hence the radiation from each Snake will be about $0.005 \times 50/1.25 \simeq 0.19$ MW over a length of 800 m.
These values should be acceptable.

\subsection{Graph}
In Fig.~\ref{fig:pol_snakes}, I plotted the data for the four energies in Table \ref{tab:pol_estimates}.
The squares are for the na{\"\i}ve asymptotic polarization and the triangles are for the
lifetime weighted average polarization.
The data at 45 GeV was computed without Snakes; the other three cases employed Snakes with $f=0.48$.
I also plotted smooth curves (solid and dashed) simply as a guide to the eye for the two datasets, respectively.
The curves are given by
\bq
P_{\rm asymp} = \frac{60}{1 + (E/90)^2} \,,\qquad
P_{\rm avg} = \frac{48}{1 + (E/80)^2} \,.
\eq

\subsection{Summary}
Snakes are not required at the $Z^0$ pole, 
and at the $W$ pair threshold one can probably perform energy calibration using non-colliding bunches, again without Snakes.
However, Snakes will probably be essential to obtain a useful degree of polarization at higher energies.
The above analysis indicates that, with several caveats,
a reasonable degree of the (lifetime averaged) polarization,
sufficient for energy calibration via resonant depolarization,
with a practical time constant, 
may be achievable even up to the highest energies in FCC-ee,
if one employs a pair of Snakes with (anti)parallel spin rotation axes.
The Snakes are {\em not} located at diametrically opposite points in the ring.
The polarization will be produced almost entirely by wigglers,
powered to have {\em opposite polarity} in the two arcs.

The above analysis clearly overestimates the degree of the asymptotic polarization.
I have treated only synchrotron sideband resonances centered on an integer.
There are obviously additional spin resonances.
However, reducing the value of the spin tune modulation index
will reduce the strengths of {\em all} the synchrotron sideband resonances,
including those centered on betatron spin resonances.
Furthermore, it should be possible to adjust the betatron tunes to optimize the polarization.
One would adjust both the integer and the fractional parts of the betatron tunes.

Of course, one could employ Snakes with non-colliding bunches and avoid the problems of the luminosity lifetime.
This would yield the data plotted as squares in Fig.~\ref{fig:pol_snakes}.
A considerable degree of polarization would be available even at the highest energies in FCC-ee, for energy calibration.
However, one could not obtain longitudinally polarized colliding beams this way.

\vfill\pagebreak
\section{Longitudinally polarized colliding beams}
The experience from HERA is invaluable for longitudinally polarized colliding beams.
The HERA team demonstrated the successful implementation of so-called `strong spin matching'
for the HERA spin rotators.
The same will have to be done in FCC-ee.
FCC-ee will have separate $e^+$ and $e^-$ rings,
hence it will be possible to have independent helicity control for the $e^+$ and $e^-$ beams.

To attain longitudinally polarized colliding beams 
first requires that there be a reasonable of degree of the asymptotic polarization
even without any spin rotators in the ring.
From my analysis above, I believe 
a useful degree of the asymptotic polarization, with a practical time constant,
should be possible at the $Z^0$ pole, using only spin rotators but no Snakes.
However, Snakes will be required at higher energies.

One obvious question, to attain longitudinal polarization at the IP,
is whether one should employ diametrically opposed Snakes ($f=\frac12$).
Then the spin tune modulation index vanishes $\sigma=0$, or is very small.
This will make the synchrotron sideband spin resonances very weak.
Of course, this would mean restructuring the ring,
i.e.~placing the Snakes in different locations in the ring,
compared to the $f=0.48$ scenario I analyzed above.
I found that nothing is gained by employing $f=0.5$.
The spin tune modulation index is already small for $f=0.48$,
and the polarization is dominated by the wigglers.
I calculated the polarization at $E=175$ GeV using $f=0.5$.
The other parameters were the same as before, 
i.e.~$Q_s=0.1$ and $\Delta\nu=0.45$ and $\mathcal{N}=28$, etc.
The numerical estimates for the asymptotic polarization and the time constant, 
etc.~are given in the last row of Table \ref{tab:pol_estimates}.
They are almost the same as in the second last row (for $f=0.48$).
Hence nothing is gained by employing diametrically opposed Snakes ($f=\frac12$).

\vfill\pagebreak

\vfill\pagebreak
\begin{table}[!htb]
\begin{tabular}{|l|r|r|r|r|r|r|}
\hline
& \multicolumn{2}{c|}{LEP} & \multicolumn{4}{c|}{TLEP/FCC-ee} \\
\hline
Circumference [km] &\multicolumn{2}{c|}{26.7} & \multicolumn{4}{c|}{100} \\
\hline
Bending radius [km] &\multicolumn{2}{c|}{3.1} & \multicolumn{4}{c|}{11} \\
\hline
Energy [GeV] \qquad & \qquad  45.4 & \qquad 104 & \qquad 45.5 & \qquad 80 & \qquad 120 & \qquad 175  \\
\hline
Energy spread [\%] \qquad & \qquad  0.07 & \qquad 0.16 & \qquad 0.06 & \qquad 0.09 & \qquad 0.14 & \qquad 0.19  \\
\hline
Synchrotron tune $Q_s$ \qquad & \qquad 0.065 & \qquad 0.083 & \qquad 0.65 & \qquad 0.21 & \qquad 0.096 &\qquad 0.10  \\
\hline
Luminosity lifetime $\tau_\ell$ [min] \qquad & \qquad  1250 & \qquad 310 & \qquad 213 & \qquad 52 & \qquad 21  &\qquad 15  \\
\hline
Polarization time $\tau_p$ [min] \qquad & \qquad  348 & \qquad 5.5 & \qquad 16238 & \qquad 966 & \qquad 127  &\qquad 19  \\
\hline
Guide field [Gauss] \qquad & \qquad  488 & \qquad 1119 & \qquad 136 & \qquad 243 & \qquad 364  &\qquad 531  \\
\hline
Tune modulation index 
$\frac{\gamma_0a}{Q_s}\,\frac{\sigma_E}{E}$
\qquad & \qquad  1.11 & \qquad 4.56 & \qquad 0.095 & \qquad 0.78 & \qquad 3.98  &\qquad 7.56  \\
\hline
\end{tabular}
\caption{\small
\label{tab:ring_params}
Parameter values at LEP and FCC-ee for various beam energies.
The values are calculated for a planar ring without wigglers.
The data for the circumference, bending radius, beam energy,
relative energy spread, synchrotron tune and luminosity lifetime
are taken from \cite{TLEPparams}.
}
\end{table}

\vfill\pagebreak
\begin{table}[!htb]
\begin{tabular}{|lr|r|r|r|r|r|r|r|r|r|r|r|r|}
\hline
&Snakes &
nc &
E [GeV] &
$\tau_p$ [min] & 
$P$ [\%] &
$P_{\rm avg}$ [\%] &
$A_U$ &
$A_p$ &
$A_{n,s}$ &
$\sigma_\varepsilon$ [\%] &
$\sigma$ &
$\mathcal{F}$ &
$\mathcal{N}$ 
\\
\hline
no &&& 45.4 & 66 & \quad 49.7 & \quad 38.0 & \quad 8.84 & \quad 220.5 & \quad 132.7 & \quad 0.19 & \quad 0.30 & \quad 1.16 & \quad 28
\\
&&& 80 & 56 & 16.7 & 7.9 & 1.49 & 4.43 & 3.058 & 0.13 & 1.11 & 9.02 & 7
\\
&& yes & 80 & 127 & 21.0 && 1.25 & 2.25 & 1.75 & 0.11 & 0.93 & 7.58 & 5
\\
\hline
yes & $f=0.48$ && 45.5 & 66 & 49.6 && 8.84 & 220.5 & 131.8 & 0.19 & 0.012 & 1.11 & 28
\\
&&& 80 & 39 & 38.5 & 22.0 & 2.44 & 18.28 & 10.4 & 0.19 & 0.065 & 1.17 & 12
\\
&&& 120 & 16 & 24.0 & 13.7 & 1.49 & 4.43 & 2.10 & 0.19 & 0.223 & 1.17 & 7
\\
&&& 175 & 3 & 11.3 & 9.4 & 1.25 & 2.25 & 0.79 & 0.23 & 0.364 & 1.25 & 5
\\
\hline
yes & $f=0.5$ && 175 & 3 & 11.4 & 9.4 & 1.25 & 2.25 & 0.75 & 0.23 & 0 & 1.13 & 5
\\
\hline
\end{tabular}
\caption{\small
\label{tab:pol_estimates}
Numerical estimates for the polarization time and asymptotic polarization in FCC-ee,
for various scenarios.
The parameters definitions are given in the text.
The caption `nc' means `non-colliding' bunches.
}
\end{table}

\vfill\pagebreak
\begin{figure}[!htb]
\centering
\includegraphics[width=0.95\textwidth]{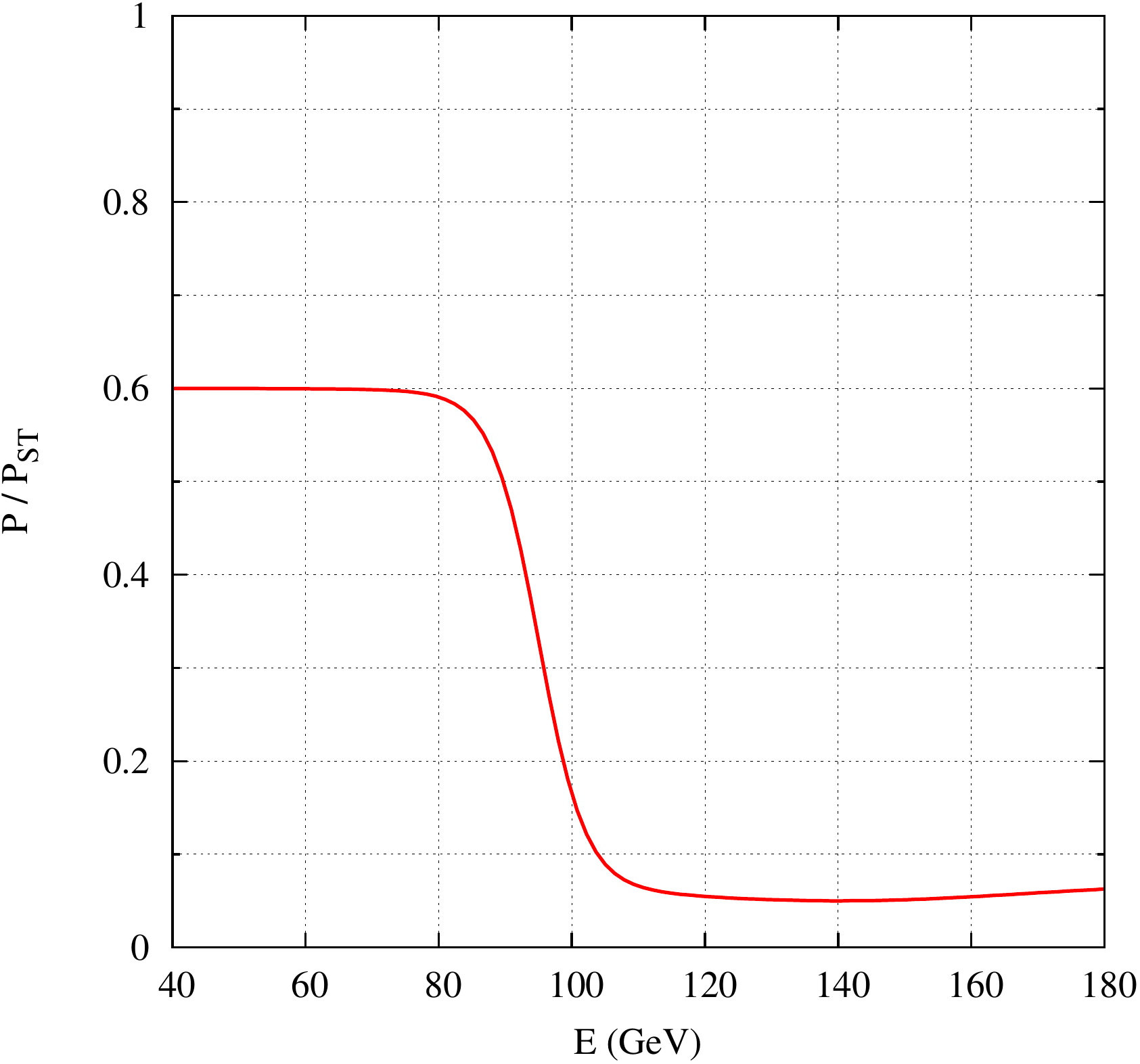}
\caption{\small
\label{fig:figulislide8}
Sketch of similar looking curve to that in slide 8 in \cite{Wienands2013}.
The plotted function is described in the text.
}
\end{figure}

\vfill\pagebreak
\begin{figure}[!htb]
\centering
\includegraphics[width=0.95\textwidth]{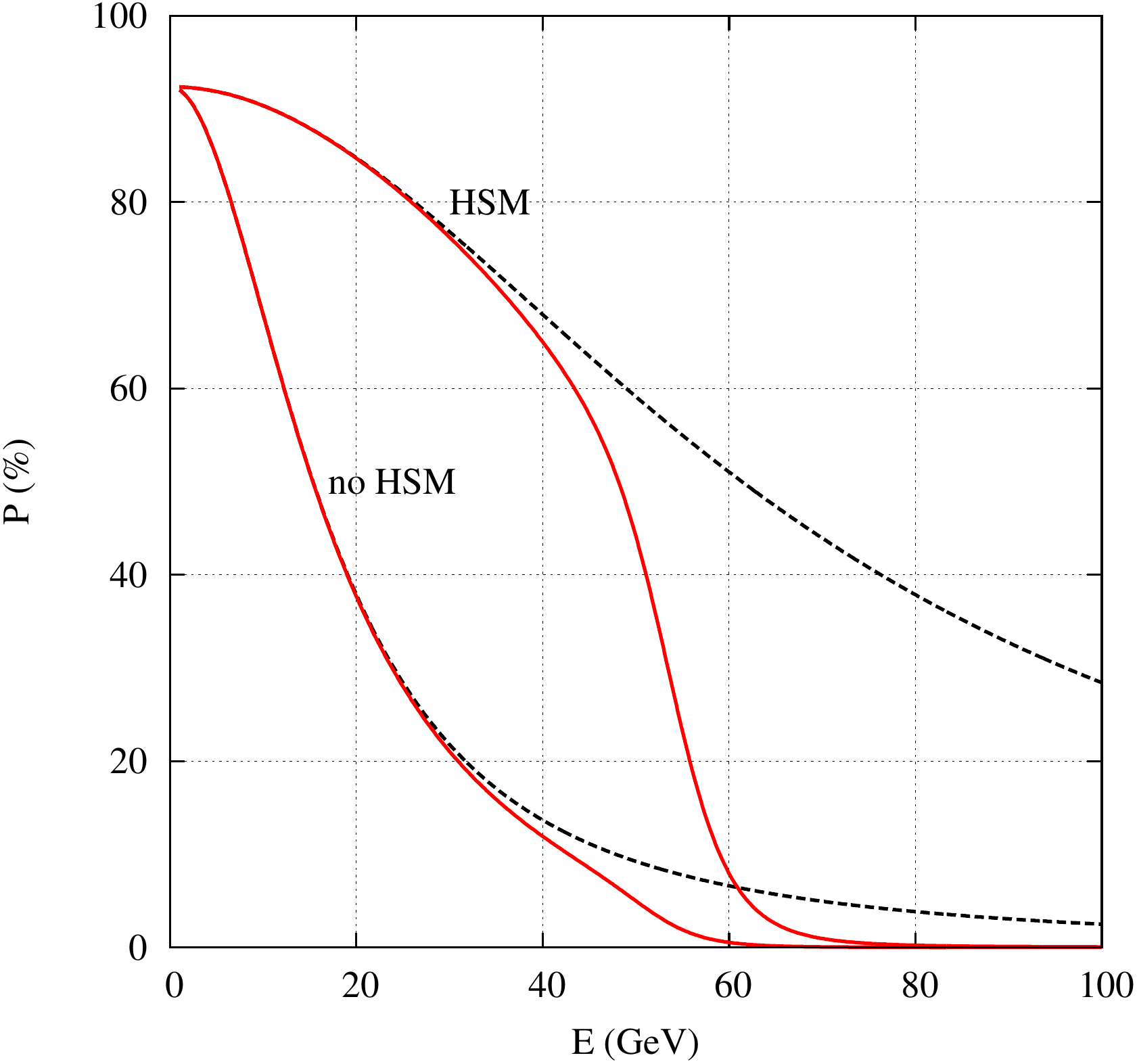}
\caption{\small
\label{fig:pol_scaling}
The asymptotic polarization computed using a scaling law, as described in the text.
The dashed curves are computed using a model of first order spin resonances only
and the solid curves are computed using a model of higher order spin resonances.
Results to simulate the cases with and without harmonic spin matching (HSM) are displayed.
}
\end{figure}

\vfill\pagebreak
\begin{figure}[!htb]
\centering
\includegraphics[width=0.95\textwidth]{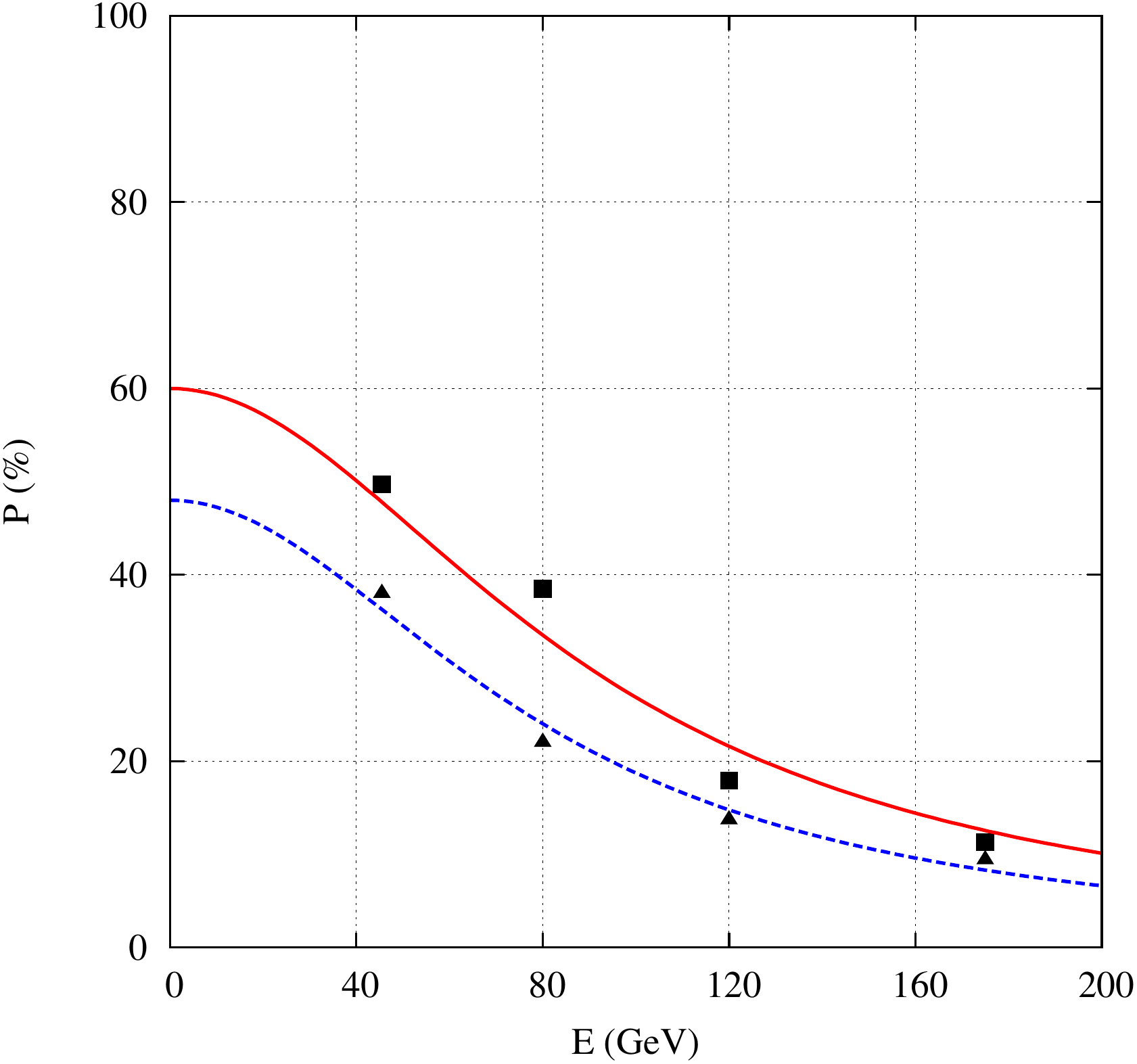}
\caption{\small
\label{fig:pol_snakes}
The estimated asymptotic polarization for a model of FCC-ee
with Snakes and wigglers, as described in the text.
The data points are from calculations tabulated in Table \ref{tab:pol_estimates}.
The curves are guides the eye, as described in the text.
}
\end{figure}

\end{document}